\documentclass[twocolumn,showpacs,preprintnumbers]{revtex4}
\usepackage{amsmath}
\usepackage{amssymb}
\usepackage{amsfonts}
\usepackage{graphicx}
\usepackage{dcolumn}
\usepackage{bm}

\input{tcilatex}

\begin{document}

\title{Quantum Replicator Dynamics}
\author{Esteban Guevara Hidalgo$^{\dag \ddag }$}
\affiliation{$^{\dag }$Departamento de F\'{\i}sica, Escuela Polit\'{e}cnica Nacional,
Quito, Ecuador\\
$^{\ddag }$SI\'{O}N, Autopista General Rumi\~{n}ahui, Urbanizaci\'{o}n Ed%
\'{e}n del Valle, Sector 5, Calle 1 y Calle A \# 79, Quito, Ecuador}

\begin{abstract}
We propose quantization relationships which would let us describe and
solution problems originated by conflicting or cooperative behaviors among
the members of a system from the point of view of quantum mechanical
interactions. The quantum analogue of the replicator dynamics is the
equation of evolution of mixed states from quantum statistical mechanics. A
system and all its members will cooperate and rearrange its states to
improve their present condition. They strive to reach the best possible
state for each of them which is also the best possible state for the whole
system. This led us to propose a quantum equilibrium in which a system is
stable only if it maximizes the welfare of the collective above the welfare
of the individual. If it is maximized the welfare of the individual above
the welfare of the collective the system gets unstable and eventually it
collapses.
\end{abstract}

\pacs{03.65.-w, 02.50.Le, 03.67.-a, 03.67.Lx}
\maketitle
\email{esteban\_guevarah@yahoo.es}

\section{Introduction}

Why quantum versions have shown be more efficient and provide more
advantages than their classical versions? Maybe quantum mechanics is a more
general theory that we had thought and it would let us explain more
correctly not only economical but also biological phenomena which have been
described before through game theory. If there is an actual connection
between game theory and quantum mechanics, we could establish quantization
relationships between our classical into quantum systems for any cases. We
could also propose a more general quantum equilibrium applicable to any
classical system but which remains inside its foundations the classical
definitions of equilibrium. The purpose of this paper is show the
relationships between the replicator dynamics and quantum mechanics. We will
propose the quantization relationships for a classical system and a new
quantum equilibrium concept from which is possible to obtain and understand
the classical equilibria.

\bigskip

Game theory \cite{1,2,3} is the study of decision making of competing agents
in some conflict situation. It tries to understand the birth and the
development of conflicting or cooperative behaviors among a group of
individuals who behave rationally and strategically according to their
personal interests. Each member in the group strive to maximize its welfare,
state, utilities or payoffs by choosing the best courses of strategies from
a cooperative or individual point of view. Game theory has been applied to
solve many problems in economics, social sciences, biology and engineering.

\bigskip

Evolutionary game theory \cite{4,5,6} does not rely on rational assumptions
but on the idea that the Darwinian process of natural selection \cite{7}
drives organisms towards the optimization of reproductive success \cite{8}.
It combines the principles of game theory, evolution and dynamical systems
to explain the distribution of different phenotypes in biological
populations. Instead of working out the optimal strategy, the different
phenotypes in a population are associated with the basic strategies that are
shaped by trial and error by a process of natural selection or learning. It
can also be used to interpret classical games from a different perspective.
Instead of directly calculating properties of a game, populations of players
using different strategies are simulated and a process similar to natural
selection is used to determine how the population evolves. This is made
through the stability analysis of differential equations and the
implications to the games \cite{9}. The central equilibrium concept of
evolutionary game theory is the notion of Evolutionary Stable Strategy
introduced by J. Smith and G. Price \cite{10,4}. An ESS is described as a
strategy which has the property that if all the members of a population
adopt it, no mutant strategy could invade the population under the influence
of natural selection. ESS are interpreted as stable results of processes of
natural selection. The natural selection process that determines how
populations playing specific strategies evolve is known as the replicator
dynamics \cite{11,5,6,9} whose stable fixed points are Nash equilibria \cite%
{2}.

\bigskip

Quantum games have proposed a new point of view for the solution of the
classical problems and dilemmas in game theory. It has been shown that
quantum games are more efficient than classical games and provide a
saturated upper bound for this efficiency \cite%
{12,13,14,15,16,17,18,19,20,21,22}. Meyer \cite{12} quantized a
coin tossing game and found out that one player could increase his
expected payoff and win with certainty by implementing a quantum
strategy against his opponent's classical strategy. Eisert
\textit{et al} \cite{13} developed a general protocol for two
player--two strategy quantum games with entanglement by quantizing
prisoner's dilemma. They found a unique Nash Equilibrium, which is
different from the classical one, and the dilemma could be solved
if the two players are allowed to use quantum strategies. This
also was extended to multiplayer games \cite{23}. Marinatto and
Weber \cite{15} extended the concept of a classical two-person
static game to the quantum domain by giving an Hilbert structure
to the space of classical strategies. They showed that the
introduction of entangled strategies in battle of the sexes game
leads to a unique solution of this game. Du \textit{et al}
\cite{17} implemented a game via nuclear magnetic resonance (NMR)
system. It was demonstrated that neither of the two players would
win the game if they play rationally, but if they adopt quantum
strategies both of them would win. Quantum games has been used to
explore unsolved problems of quantum information \cite{24} and in
the production of algorithms for quantum computers \cite{25}.
Quantum communication can be considered as a game where the
objective is maximize effective communication. Also distributed
computing, cryptography, watermarking and information hiding tasks
can be modeled as games \cite{26,27,28,29,30,31,32}. Piotrowski
and Sladkowski have modeled markets, auctions and bargaining
assuming traders can use quantum protocols \cite{33,34}. In the
new quantum market games, transactions are described in terms of
projective operations acting on Hilbert spaces of strategies of
traders. A quantum strategy represents a superposition of trading
actions and can achieve outcomes not realizable by classical means
\cite{35}. Furthermore, quantum mechanics has features that can be
used to model aspects of market behavior. For example, traders
observe the actions of other players and adjust their actions
\cite{34} and the maximal capital flow at a given price
corresponds to entanglement between buyers and sellers \cite{33}.
Nature may be playing quantum survival games at the molecular
level \cite{36,37}. It could lead us to describe many
of the life processes through quantum mechanics like Gogonea and Merz \cite%
{38} on protein molecules. Game theory and quantum game theory offer
interesting and powerful tools and their results will probably find their
applications in computation, complex system analysis and cognition sciences
\cite{39,40,41}.

\section{Replicator Dynamics \& EGT}

Payoffs in biological games are in terms of fitness a measure of
reproductive success \cite{8}. Strategies are considered to be inherited
programs for any conceivable situation which control the individual's
behavior. The members of a population interact in game situations and the
joint action of mutation and selection\ replaces strategies by others with a
higher reproductive success. It is less important to know which member plays
which strategy within a population but it is important to know the relative
frequency of actions, it means the probability of playing any strategy.

\bigskip

A\ \textit{symmetric two - person game }$G=(S,E)$ consists of a finite
nonempty pure strategy set $S$ and a payoff function $E$ which assigns a
real number to the pair $(s_{i},s_{j})$. $E(s_{i},s_{j})$ is the payoff
obtained by a player who plays the strategy $s_{i}$\ against an opponent who
plays the strategy $s_{j}$. A \textit{mixed strategy }$x$ is a probability
distribution over $S$.

A\textit{\ best reply} to $q$ is a strategy $p$ which maximizes $E(p,q)$. An
\textit{equilibrium point} is a pair $(p,q)$ with the property that $p$ and $%
q$ are best replies to each other. A strategy $r$ is a \textit{strict best
reply} to a strategy $q$ if it is the only best reply to $q$. A strict best
reply must be a pure strategy. An equilibrium point $(p,q)$ is called strict
if $p$ and $q$ are strict best replies to each other. A best reply to $p$
which is different from $p$ is called \textit{alternative best reply} to $p$.

A \textit{Nash equilibrium }(NE) is a set of strategies, one for each
player, such that no player has an incentive to unilaterally change his
action. Players are in equilibrium if a change in strategies by any one of
them would lead that player to earn less than if he remained with his
current strategy. A \textit{Nash equilibrium} satisfies the following
condition%
\begin{equation}
E(p,p)\geq E(r,p)\text{.}  \label{1}
\end{equation}%
A player can not increase his payoff if he decides to play the strategy $r$
instead of $p.$

\bigskip

Each agent in a n-player game where the i$^{\text{th}}$ player has
as strategy space $S_{i}$ is modeled by a population of players
which have to be partitioned into groups. Individuals in the same
group would all play the same strategy. Randomly we make play the
members of the subpopulations against each other. The
subpopulations that perform the best will grow and those that do
not will shrink and eventually will vanish. The process of natural
selection assures survival of the best players at the expense of
the others. A population equilibrium occurs when the population
shares are such that the expected payoffs for all strategies are
equal.

\bigskip

Consider a large population in which a two person game $G=(S,E)$ is played
by randomly matched pairs of animals generation after generation. Let $p$ be
the strategy played by the vast majority of the population, and let $r$ be
the strategy of a mutant present in small frequency. Both $p$ and $r$ can be
pure or mixed. An \textit{evolutionary stable strategy} (ESS) $p$ of a
symmetric two - person game $G=(S,E)$ is a pure or mixed strategy for $G$
which satisfies the following two conditions%
\begin{gather}
E(p,p)>E(r,p)\text{,}  \notag \\
\text{If }E(p,p)=E(r,p)\text{ then }E(p,r)>E(r,r)\text{.}  \label{2}
\end{gather}%
Since the stability condition only concerns to alternative best replies, $p$
is always evolutionarily stable if $(p,p)$ is an strict equilibrium point.
An ESS is also a Nash equilibrium since it is the best reply to itself and
the game is symmetric. The set of all the strategies that are ESS is a
subset of the NE of the game. A population which plays an ESS can withstand
an invasion by a small group of mutants playing a different strategy. It
means that if a few individuals which play a different strategy are
introduced into a population in an ESS, the evolutionarily selection process
would eventually eliminate the invaders.

\bigskip

The natural selection process that determines how populations playing
specific strategies evolve is known as the \textit{replicator dynamics}.%
\textit{\ }It describes the evolution of a polymorphic state in a population
represented by a mixed strategy $x$ for $G$ whose members are involved in a
conflict described by a symmetric two - person game $G=(S,E)$. The
probability assigned to a pure strategy $s$ is denoted by $x(s)$. If $%
s_{i},i=1,...,n\in S$ are the pure strategies available to a player, then
that player's strategy will be denoted by the column vector $x$ where $%
x_{i}\in \left[ 0,1\right] $ and $\tsum_{i=1}^{n}x_{i}=1$ in which the $%
i^{th}$ component of $x$ gives the probability of playing strategy $s_{i}$.
The $i^{th}$ component of $x$ also is interpreted as the relative frequency
of \ individuals using strategy $s_{i}$. Playing a pure strategy $s_{j}$ is
represented by the vector $x$ whose $j^{th}$ component is $1$, and all the
other components are $0$. The fitness function is given by $E=f_{i}(x),$ $%
i=1,...,n$ and specifies how successful each subpopulation is. The fitness
function must be defined for each component of $x$. The fitness for $x_{i}$
is the expected utility of playing strategy $s_{i}$ against a player with a
mixed strategy defined by the vector $x$ and is given by%
\begin{equation}
f_{i}(x)=(Ax)_{i}\text{,}  \label{3}
\end{equation}%
where $A$ is the payoff matrix and the subscript $i=1,...,n$ denotes the i$^{%
\text{th}}$ component of the matrix-vector product. Or also%
\begin{equation}
f_{i}(x)=\sum_{j=1}^{n}a_{ij}x_{j}\text{,}  \label{4}
\end{equation}%
where $a_{ij}$ are the elements of the payoff matrix $A$. The average
fitness of the population is given by%
\begin{gather}
\left\langle f(x)\right\rangle =x^{T}Ax\text{,}  \label{5} \\
\left\langle f(x)\right\rangle =\sum_{i=1}^{n}x_{i}f_{i}(x)\text{,}
\label{6} \\
\left\langle f(x)\right\rangle =\sum_{k,l=1}^{n}a_{kl}x_{k}x_{l}\text{.}
\label{7}
\end{gather}%
The superscript $T$ denotes transpose.

\bigskip

The evolution of relative frequencies in a population is described by the
replicator dynamics%
\begin{equation}
\frac{dx_{i}(t)}{dt}=\left[ f_{i}(x)-\left\langle f(x)\right\rangle \right]
x_{i}(t)\text{.}  \label{8}
\end{equation}%
By replacing relations (\ref{3}) and (\ref{5}) in equation (\ref{8}) and by
replacing relations (\ref{4}) and (\ref{7}) we can obtain two different ways
of representing the replicator dynamics%
\begin{gather}
\frac{dx_{i}(t)}{dt}=\left[ (Ax)_{i}-x^{T}Ax\right] x_{i}(t)\text{,}
\label{9} \\
\frac{dx_{i}(t)}{dt}=\left[ \sum_{j=1}^{n}a_{ij}x_{j}-%
\sum_{k,l=1}^{n}a_{kl}x_{k}x_{l}\right] x_{i}(t)\text{.}  \label{10}
\end{gather}%
The stable fixed points of the replicator dynamics are Nash equilibria. It
is important to note that the fixed points of a system do not change in the
time. It means that if a population reaches a state which is a Nash
equilibrium, it will remain there. There are different versions of these
equations \cite{5,6,9,11} depending on the evolutionary model used. The
replicator dynamics rewards strategies that outperform the average by
increasing their frequency, and penalizes poorly performing strategies by
decreasing their frequency.

\bigskip

In a symmetric game payoff matrices and actions are identical for
both agents. These games can be modeled by a single population of
individuals playing against each other. When the game being played
is asymmetric, a different population of players must be used to
simulate each agent. For a two-player asymmetric normal game each
player will have a distinct payoff matrix $A$ and $B$. The
strategy vector for player one is represented by $x\
$and for player two is represented by $y$. Player one has $n$ strategies $%
s_{1i},i=1,...,n\in S_{1}$\ \ and player two has $m$ strategies $%
s_{2j},j=1,...,m\in S_{2}$. The fitness for a player who plays the strategy\
$s_{1i}$\ will be $f_{1i}=(Ay)_{i}$\ and the average fitness of the first
population will be $\left\langle f_{1}\right\rangle =x^{T}Ay$. The fitness
for a player who plays the strategy\ $s_{2i}$\ will be $f_{2i}=(Bx)_{i}$\
and the average fitness of the second population\ will be\ $\left\langle
f_{2}\right\rangle =y^{T}Bx$. The evolution of this game would be described
for the next equations system%
\begin{gather}
\frac{dx_{i}(t)}{dt}=\left[ (Ay)_{1}-x^{T}Ay\right] x_{i}(t)\text{,}  \notag
\\
\frac{dy_{i}(t)}{dt}=\left[ (Bx)_{1}-y^{T}Bx\right] y_{i}(t)\text{.}
\label{11}
\end{gather}%
The genetic code is the relationship between the sequence of the bases in
the DNA and the sequence of amino acids in proteins. Recent work \cite{42}
about evolvability of the genetic code suggests that the code is shaped by
natural selection.

\section{Replicator Dynamics Matrix form}

As we saw the replicator dynamics is a differential equation where $x$ is a
column vector. Obviously, the matrix $U=(Ax)_{i}-x^{T}Ax$ has to be diagonal
and its elements are given by%
\begin{equation}
u_{ii}=\sum_{j=1}^{n}a_{ij}x_{j}-\sum_{k,l=1}^{n}a_{kl}x_{k}x_{l}\text{.}
\label{12}
\end{equation}%
The replicator dynamics can be expressed as%
\begin{equation}
\frac{dx}{dt}=Ux\text{.}  \label{13}
\end{equation}%
By multiplying each element of the vector $x$ by its corresponding $%
(x_{i})^{-1/2}$ in both parts of equation (\ref{10}) or (\ref{13}), we can
get
\begin{equation}
v=U\hat{x}\text{,}  \label{14}
\end{equation}%
where $v$ and $\hat{x}$ are column vectors with elements $v_{i}=\frac{1}{%
(x_{i})^{1/2}}\frac{dx_{i}}{dt}$ and $\hat{x}_{i}=(x_{i})^{1/2}$
respectively. Lets multiply equation (\ref{14}) by $\hat{x}^{T}$ and lets
define the matrix%
\begin{equation}
G=\frac{1}{2}v\hat{x}^{T}\text{,}  \label{15}
\end{equation}%
where $g_{ij}=\frac{1}{2}\frac{(x_{j})^{1/2}}{(x_{i})^{1/2}}\frac{dx_{i}}{dt}
$ are the elements of $G$. The matrix $G$ is also equal to%
\begin{equation}
G=\frac{1}{2}U\hat{x}\hat{x}^{T}  \label{16}
\end{equation}%
with

$(U\hat{x}\hat{x}^{T})_{ij}=\left(
\sum_{k=1}^{n}a_{ik}x_{k}-\sum_{k,l=1}^{n}a_{kl}x_{k}x_{l}\right)
(x_{i}x_{j})^{1/2}$. Lets find $G^{T}$%
\begin{gather}
G^{T}=\frac{1}{2}(v\hat{x}^{T})^{T}\text{,}  \notag \\
G^{T}=\frac{1}{2}\hat{x}v^{T}\text{,}  \label{17}
\end{gather}%
where $g_{ij}^{T}=\frac{1}{2}\frac{(x_{i})^{1/2}}{(x_{j})^{1/2}}\frac{dx_{j}%
}{dt}$ are the elements of $G^{T}$. The matrix $G^{T}$ is also equal to%
\begin{gather}
G^{T}=\frac{1}{2}(U\hat{x}\hat{x}^{T})^{T}\text{,}  \notag \\
G^{T}=\frac{1}{2}\hat{x}\hat{x}^{T}U^{T}  \label{18}
\end{gather}%
with

$(\hat{x}\hat{x}^{T}U^{T})_{ij}=(x_{j}x_{i})^{1/2}\left(
\sum_{k=1}^{n}a_{jk}x_{k}-\sum_{k,l=1}^{n}a_{kl}x_{k}x_{l}\right) $. Lets
define%
\begin{equation}
\frac{dX}{dt}=G+G^{T}  \label{19}
\end{equation}%
so that the matrix $X$ has as elements%
\begin{equation}
x_{ij}=\left( x_{i}x_{j}\right) ^{1/2}\text{.}  \label{20}
\end{equation}%
The time evolution of $X$ is also equal to%
\begin{equation}
G+G^{T}=\frac{1}{2}(U\hat{x}\hat{x}^{T}+\hat{x}\hat{x}^{T}U^{T})  \label{21}
\end{equation}%
with%
\begin{eqnarray}
\left( G+G^{T}\right) _{ij} &=&\frac{1}{2}%
\sum_{k=1}^{n}a_{ik}x_{k}(x_{i}x_{j})^{1/2}  \notag \\
&&+\frac{1}{2}\sum_{k=1}^{n}a_{jk}x_{k}(x_{j}x_{i})^{1/2}  \notag \\
&&-\sum_{k,l=1}^{n}a_{kl}x_{k}x_{l}(x_{i}x_{j})^{1/2}\text{.}  \label{22}
\end{eqnarray}%
Lets call%
\begin{gather}
\left( G_{1}\right) _{ij}=\frac{1}{2}%
\sum_{k=1}^{n}a_{ik}x_{k}(x_{i}x_{j})^{1/2}\text{,}  \label{23} \\
\left( G_{2}\right) _{ij}=\frac{1}{2}%
\sum_{k=1}^{n}a_{jk}x_{k}(x_{j}x_{i})^{1/2}\text{,}  \label{24} \\
\left( G_{3}\right) _{ij}=\sum_{k,l=1}^{n}a_{kl}x_{k}x_{l}(x_{i}x_{j})^{1/2}
\label{25}
\end{gather}%
the elements of the matrixes $G_{1}$, $G_{2}$ and $G_{3}$ that compose by
adding the matrix $\left( G+G^{T}\right) $. We can factorize $G_{1}$ in two
matrixes $Q$ and $X$. It means%
\begin{equation}
G_{1}=QX\text{,}  \label{26}
\end{equation}%
where $Q$ is a diagonal matrix and has as elements $q_{ii}=\frac{1}{2}%
\sum_{k=1}^{n}a_{ik}x_{k}$. In the same way%
\begin{equation}
G_{2}=XQ\text{.}  \label{27}
\end{equation}%
The matrix $G_{3}$ can also be represented as $\left( G_{3}\right)
_{ij}=\sum_{l=1}^{n}(x_{i}x_{l})^{1/2}%
\sum_{k=1}^{n}a_{kl}x_{k}(x_{l}x_{j})^{1/2}$ and it can be factorized into%
\begin{equation}
G_{3}=2XQX\text{.}  \label{28}
\end{equation}%
It is easy to show that $X^{2}=X$ so that we can write the equation (\ref{19}%
) like%
\begin{equation}
\frac{dX}{dt}=QXX+XXQ-2XQX  \label{29}
\end{equation}%
and finally, by grouping into commutators%
\begin{gather}
\frac{dX}{dt}=\left[ \left[ Q,X\right] ,X\right] \text{,}  \label{30} \\
\frac{dX(t)}{dt}=\left[ \Lambda (t),X(t)\right] \text{,}  \label{31}
\end{gather}%
where $\Lambda =\left[ Q,X\right] $.

The elements of $\Lambda $ are given \ by $(\Lambda )_{ij}=\frac{1}{2}\left[
\left( \sum_{k=1}^{n}a_{ik}x_{k}\right)
(x_{i}x_{j})^{1/2}-(x_{j}x_{i})^{1/2}\left( \sum_{k=1}^{n}a_{jk}x_{k}\right) %
\right] $. Equations (\ref{30}) and (\ref{31}) can be considered as a
generalization of the replicator dynamics as was seen in equations (\ref{8}%
), (\ref{9}) and (\ref{10}). We could call $X$ the relative frequency matrix
and its evolution is described by equation (\ref{31}). Matrix $X$ has the
following properties

\begin{enumerate}
\item[a)] $Tr(X)=1$,

\item[b)] $X^{2}=X$,

\item[c)] $X^{T}=X$.
\end{enumerate}

\bigskip

It is easy to realize that each component of this matrix will evolve
following the replicator dynamics as was seen in equations (\ref{8}), (\ref%
{9}) and (\ref{10}) so that we could call equation (\ref{31}) the matrix
form of the replicator dynamics. If we take $\Theta =\left[ \Lambda ,X\right]
$ equation (\ref{31}) becomes into%
\begin{equation}
\frac{dX}{dt}=\Theta \text{,}  \label{32}
\end{equation}%
where the elements of $\ \Theta $ are given by
\begin{eqnarray}
\left( \Theta \right) _{ij} &=&\frac{1}{2}%
\sum_{k=1}^{n}a_{ik}x_{k}(x_{i}x_{j})^{1/2}  \notag \\
&&+\frac{1}{2}\sum_{k=1}^{n}a_{jk}x_{k}(x_{j}x_{i})^{1/2}  \notag \\
&&-\sum_{k,l=1}^{n}a_{lk}x_{k}x_{l}(x_{i}x_{j})^{1/2}\text{.}  \label{33}
\end{eqnarray}%
The fact that we can calculate the evolution of interference effects between
relative frequencies like $(x_{i}x_{j})^{1/2}$ is the advantage of using
equation (\ref{31}) besides equation (\ref{10}).

\section{Mixed and Pure States in Quantum Mechanics}

In classical mechanics we can precisely specify the state of a system by one
point in its phase space. Its trajectory through the phase space describes
the time evolution of the system and this evolution follows Newton's laws or
Hamilton equations. In quantum mechanics we can precisely describe the state
of a system by specifying its state vector $\left\vert \Psi (t)\right\rangle
$ or its wave function $\Psi (r,t)$ and the evolution of the system is given
by Schr\"{o}dinger equation. However, in most cases when the information
about the system is incomplete the state of a system is not perfectly
defined for which we have to describe our system in terms of probabilities
\cite{43,44}.

\bigskip

An \textit{ensemble} is a collection of identically prepared physical
systems. When each member of the ensemble is characterized by the same state
vector $\left\vert \Psi (t)\right\rangle $ it is called \textit{pure ensemble%
}. If each member has a probability $p_{i}$ of being in the state $%
\left\vert \Psi _{i}(t)\right\rangle $ we have a \textit{mixed ensemble}.
The state of a system $\left\vert \Psi _{i}(t)\right\rangle $ $=\left\vert
\Psi _{k}^{(i)}(t)\right\rangle $ is a column vector $\left\vert \Psi
_{k}^{(i)}(t)\right\rangle =(\Psi _{1}^{(i)}(t),\Psi _{2}^{(i)}(t),...,\Psi
_{n}^{(i)}(t))^{T},k=1,...,n$.

\bigskip

For the case of \textit{pure ensemble }the expected value of measure certain
observable described by the operator $A$ in the state $\left\vert \Psi
(t)\right\rangle $ at the instant $t$ is given by%
\begin{gather}
\left\langle A\right\rangle =\left\langle \Psi (t)\right\vert A\left\vert
\Psi (t)\right\rangle =\sum_{i,j=1}^{n}\left\langle j\left\vert \Psi
(t)\right. \right\rangle a_{ij}\left\langle \Psi (t)\left\vert i\right.
\right\rangle \text{,}  \notag \\
\left\langle A\right\rangle =\sum_{i,j=1}^{n}a_{ij}c_{i}^{\ast }(t)c_{j}(t)%
\text{,}  \label{34}
\end{gather}%
where $a_{ij}$ are the elements of the matrix that represents the observable
$A.$ The terms $c_{i}^{\ast }(t)=\left\langle \Psi (t)\left\vert i\right.
\right\rangle $ and $c_{j}(t)=\left\langle j\left\vert \Psi (t)\right.
\right\rangle $ are the elements of certain operator $\rho (t)$%
\begin{equation}
c_{i}^{\ast }(t)c_{j}(t)=\left\langle j\left\vert \Psi (t)\right.
\right\rangle \left\langle \Psi (t)\left\vert i\right. \right\rangle
=\left\langle j\right\vert \rho (t)\left\vert i\right\rangle   \label{35}
\end{equation}%
defined as
\begin{equation}
\rho (t)=\left\vert \Psi (t)\right\rangle \left\langle \Psi (t)\right\vert
\text{.}  \label{36}
\end{equation}%
The density operator $\rho (t)$ for a pure ensemble in a state $\left\vert
\Psi (t)\right\rangle $ satisfies the following properties

\begin{enumerate}
\item[a)] $\rho ^{+}(t)=\rho (t)$,

\item[b)] $\rho ^{2}(t)=\rho (t)$,

\item[c)] $Tr\rho ^{2}(t)=1$.
\end{enumerate}

The superscript $^{+}$ denotes Hermitian. We can make all the predictions
made for $\left\vert \Psi (t)\right\rangle $ in function of $\rho (t)$. The
sum of the diagonal elements of the density matrix is one%
\begin{equation}
\sum_{n}\left\vert c_{n}(t)\right\vert ^{2}=Tr\rho (t)=1\text{.}  \label{37}
\end{equation}%
The mean value of an observable $A$ is calculated using%
\begin{equation}
\left\langle A\right\rangle =Tr\left\{ \rho (t)A\right\} \text{.}  \label{38}
\end{equation}%
The time evolution of the density operator is given by the following equation%
\begin{equation}
i\hbar \frac{d\rho (t)}{dt}=\left[ H(t),\rho (t)\right] \text{.}  \label{39}
\end{equation}

\bigskip

Each member of a \textit{mixed ensemble }is a pure state and its evolution
is given by Schr\"{o}dinger equation. The probabilities for each state are
constrained to satisfy the normalization condition%
\begin{equation}
\sum_{i=1}^{n}p_{i}=1\text{,}  \label{40}
\end{equation}%
where $0\leq p_{1,}p_{2},...,p_{n}\leq 1$. Suppose we make a measurement on
a mixed ensemble of some observable $A$. The \textit{ensemble average} of $A$
is defined by the average of the expected values measured in each member of
the ensemble described by $\left\vert \Psi _{i}(t)\right\rangle $\ and with
probability $p_{i}$%
\begin{equation}
\left\langle A\right\rangle _{\rho }=\sum_{i=1}^{n}p_{i}\left\langle \Psi
_{i}(t)\right\vert A\left\vert \Psi _{i}(t)\right\rangle   \label{41}
\end{equation}%
it means $\left\langle A\right\rangle _{\rho }=p_{1}\left\langle
A\right\rangle _{1}+p_{2}\left\langle A\right\rangle
_{2}+...+p_{n}\left\langle A\right\rangle _{n}.$%
\begin{gather}
\left\langle A\right\rangle _{\rho }=\sum_{i=1}^{n}p_{i}\left\langle \Psi
_{i}(t)\right\vert A\left\vert \Psi _{i}(t)\right\rangle \text{,}  \notag \\
\left\langle A\right\rangle _{\rho }=\sum_{i,j,k=1}^{n}p_{i}\left\langle
\Psi _{i}(t)\left\vert j\right. \right\rangle \left\langle j\right\vert
A\left\vert k\right\rangle \left\langle k\left\vert \Psi _{i}(t)\right.
\right\rangle \text{,}  \notag \\
\left\langle A\right\rangle _{\rho
}=\sum_{i,j,k=1}^{n}p_{i}a_{jk}c_{j}^{(i)\ast }(t)c_{k}^{(i)}(t)\text{,}
\label{42}
\end{gather}%
where $a_{jk}$ are the elements of the matrix that represents the observable
$A$. The terms $c_{k}^{(i)}(t)=\left\langle k\left\vert \Psi _{i}(t)\right.
\right\rangle $ and $c_{j}^{(i)\ast }(t)=\left\langle \Psi _{i}(t)\left\vert
j\right. \right\rangle $ are the elements of certain operator $\rho (t)$%
\begin{gather}
\sum_{i=1}^{n}p_{i}c_{j}^{(i)\ast
}(t)c_{k}^{(i)}(t)=\sum_{i=1}^{n}\left\langle k\left\vert \Psi
_{i}(t)\right. \right\rangle p_{i}\left\langle \Psi _{i}(t)\left\vert
j\right. \right\rangle \text{,}  \notag \\
\sum_{i=1}^{n}p_{i}c_{j}^{(i)\ast }(t)c_{k}^{(i)}(t)=\left\langle
k\right\vert \rho (t)\left\vert j\right\rangle   \label{43}
\end{gather}%
now defined as%
\begin{equation}
\rho (t)=\sum_{i=1}^{n}p_{i}\left\vert \Psi _{i}(t)\right\rangle
\left\langle \Psi _{i}(t)\right\vert \text{.}  \label{44}
\end{equation}%
The density operator contains all the physically significant information we
can possibly obtain about the ensemble in question. Any two ensembles that
produce the same density operator are physically indistinguishable. From
equation (\ref{42}) we have%
\begin{equation}
\left\langle A\right\rangle =Tr\left\{ \rho (t)A\right\} \text{.}  \label{45}
\end{equation}%
The density operator for a mixed ensemble satisfies the next properties

\begin{enumerate}
\item[a)] $\rho $ is Hermitian,

\item[b)] $Tr\rho (t)=1$,

\item[c)] $\rho ^{2}(t)\leqslant \rho (t)$,

\item[d)] $Tr\rho ^{2}(t)\leqslant 1$.
\end{enumerate}

A pure state is specified by $p_{i}=1$ for some $\left\vert \Psi
_{i}(t)\right\rangle ,i=1,...,n$ and the density operator $\rho (t)$ is
represented by a matrix with all its elements equal to zero except one 1 on
the diagonal.

\bigskip

The diagonal elements $\rho _{nn}$ of \ the density operator $\rho (t)$
represents the average probability of finding the system in the state $%
\left\vert n\right\rangle $%
\begin{gather}
\rho _{nn}=\left\langle n\right\vert \rho (t)\left\vert n\right\rangle
=\sum_{i=1}^{n}\left\langle n\left\vert \Psi _{i}(t)\right. \right\rangle
p_{i}\left\langle \Psi _{i}(t)\left\vert n\right. \right\rangle \text{,}
\notag \\
\rho _{nn}=\sum_{i=1}^{n}p_{i}\left\vert c_{n}^{(i)}\right\vert ^{2}\text{,}
\label{46}
\end{gather}%
where $c_{n}^{(i)}=\left\langle n\left\vert \Psi i(t)\right. \right\rangle $
and $\left\vert c_{n}^{(i)}\right\vert ^{2}\in
\mathbb{R}
^{+}$. If the state of the system is $\left\vert \Psi _{i}(t)\right\rangle $%
, $\left\vert c_{n}^{(i)}\right\vert ^{2}$ is the probability of finding, in
a measurement, this system in the state $\left\vert n\right\rangle $. The
diagonal elements $\rho _{nn}$ are zero if and only if all $\left\vert
c_{n}^{(i)}\right\vert ^{2}$ are zero. The non-diagonal elements $\rho _{np}$
expresses the interference effects between the states $\left\vert
n\right\rangle $ and $\left\vert p\right\rangle $ which can appear when the
state $\left\vert \Psi _{i}\right\rangle $ is a coherent linear
superposition of these states%
\begin{gather}
\rho _{np}=\left\langle n\right\vert \rho (t)\left\vert p\right\rangle
=\sum_{i=1}^{n}\left\langle n\left\vert \Psi _{i}(t)\right. \right\rangle
p_{i}\left\langle \Psi _{i}(t)\left\vert p\right. \right\rangle \text{,}
\notag \\
\rho _{np}=\sum_{i=1}^{n}p_{i}\text{ }c_{n}^{(i)}(t)c_{p}^{(i)\ast }(t)
\label{47}
\end{gather}%
with $c_{n}^{(i)}(t)=\left\langle n\left\vert \Psi _{i}(t)\right.
\right\rangle $, $c_{p}^{(i)\ast }(t)=$ $\left\langle \Psi _{i}(t)\left\vert
p\right. \right\rangle $ and $c_{n}^{(i)}c_{p}^{(i)\ast }\in
\mathbb{C}
$. If $\rho _{np}=0$, it means that the average has canceled out
any interference effects between $\left\vert n\right\rangle $ and
$\left\vert p\right\rangle $ but if it is different from zero
subsists certain coherence between these states.

\bigskip

The time evolution of the density operator that describes a mixed or pure
ensemble is given by%
\begin{equation}
i\hbar \frac{d\rho (t)}{dt}=\left[ H(t),\rho (t)\right]  \label{48}
\end{equation}%
which is the quantum analogue of Liouville's theorem from classical
mechanics and a generalization of Schr\"{o}dinger equation.

\section{Density Operator \& Quantum Statistical Mechanics}

Now lets review briefly the connection between the density operator and
quantum statistical mechanics \cite{44}. The basic assumption we make to
obtain density operator from the study of an ensemble in thermal equilibrium
is that nature tends to maximize $\sigma $ subject to the constraint that
the ensemble average of the Hamiltonian has a certain prescribed value. The
quantity $\sigma $ is defined by%
\begin{equation}
\sigma =-Tr(\rho \ln \rho )  \label{49}
\end{equation}%
which can be regarded as a quantitative measure of disorder.

\bigskip

It is easy to show that for a completely random ensemble\ in where the
density matrix diagonalized looks like%
\begin{equation}
\rho =\frac{1}{N}%
\begin{pmatrix}
1 &  &  &  &  &  & 0 \\
& 1 &  &  &  &  &  \\
&  & 1 &  &  &  &  \\
&  &  & ... &  &  &  \\
&  &  &  & 1 &  &  \\
&  &  &  &  & 1 &  \\
0 &  &  &  &  &  & 1%
\end{pmatrix}
\label{50}
\end{equation}%
the quantity $\sigma $ takes the value%
\begin{equation}
\sigma =\ln N  \label{51}
\end{equation}%
and in contrast, for a pure ensemble with a density matrix diagonalized
similar to%
\begin{equation}
\rho =%
\begin{pmatrix}
0 &  &  &  &  &  & 0 \\
& 0 &  &  &  &  &  \\
&  & 1 &  &  &  &  \\
&  &  & 0 &  &  &  \\
&  &  &  & ... &  &  \\
&  &  &  &  & 0 &  \\
0 &  &  &  &  &  & 0%
\end{pmatrix}
\label{52}
\end{equation}%
the quantity $\sigma $ takes the value%
\begin{equation}
\sigma =0\text{.}  \label{53}
\end{equation}%
A pure ensemble is an ensemble with a maximum amount of order because all
members are characterized by the same quantum mechanical state ket. By the
other hand, a completely random ensemble in which all quantum mechanical
states are equally likely $\sigma $ takes its maximum possible value subject
to the normalization condition%
\begin{equation}
\sum_{k}\rho _{kk}=1\text{.}  \label{54}
\end{equation}%
Our quantity $\sigma $ is related to the definition of entropy $S$ from
quantum statistical mechanics via%
\begin{equation}
S=k\sigma \text{,}  \label{55}
\end{equation}%
where $k$ is the Boltzmann constant. Once that $\sigma $ is a maximum and
thermal equilibrium is established, we expect%
\begin{equation}
\frac{d\rho }{dt}=0\text{.}  \label{56}
\end{equation}%
We maximize $\sigma $ by requiring that%
\begin{equation}
\delta \sigma =0  \label{57}
\end{equation}%
subject to the constrains%
\begin{equation}
\delta \left\langle H\right\rangle =\sum_{k}\delta \rho _{kk}E_{k}=0
\label{58}
\end{equation}%
and%
\begin{equation}
\delta (Tr(\rho ))=\sum_{k}\delta \rho _{kk}=0\text{.}  \label{59}
\end{equation}%
By the use of Lagrange multipliers we can accomplish these and we can obtain%
\begin{equation}
\sum_{k}\delta \rho _{kk}\left[ \left( \ln \rho _{kk}+1\right) +\beta
E_{k}+\gamma \right] =0  \label{60}
\end{equation}%
which for an arbitrary variation is possibly only if%
\begin{equation}
\rho _{kk}=\exp (-\beta E_{k}-\gamma -1)\text{,}  \label{61}
\end{equation}%
where we can eliminate $\gamma $ using the normalization condition and our
final result is%
\begin{equation}
\rho _{kk}=\frac{\exp (-\beta E_{k})}{\sum\limits_{l}\exp (-\beta E_{l})}%
\text{.}  \label{62}
\end{equation}%
The denominator of equation (\ref{62}) is the partition function%
\begin{equation}
Z=\sum\limits_{l}\exp (-\beta E_{l})=Tr(e^{-\beta H})\text{.}  \label{63}
\end{equation}%
Given $\rho _{kk}$ in the energy basis we can write the density operator as%
\begin{equation}
\rho =\frac{e^{-\beta H}}{Z}\text{.}  \label{64}
\end{equation}%
The operator $\rho $ is the quantum mechanical analogue of the equilibrium
density of points in the phase space for an canonical ensemble.

\bigskip

The diagonal elements of the density operator in the $\left\{ \left\vert
n\right\rangle \right\} $ basis of eigenvectors of $H$ are given by
\begin{equation}
\rho _{nn}=\frac{\left\langle n\right\vert e^{-\beta H}\left\vert
n\right\rangle }{Z}=\frac{e^{-\beta E_{n}}}{Z}  \label{65}
\end{equation}%
and the non diagonal elements are equal to%
\begin{equation}
\rho _{np}=\frac{\left\langle n\right\vert e^{-\beta H}\left\vert
p\right\rangle }{Z}=\frac{e^{-\beta E_{p}}}{Z}\left\langle n\left\vert
p\right. \right\rangle =0\text{.}  \label{66}
\end{equation}%
At thermodynamic equilibrium, the populations of the stationary states are
exponentially decreasing functions of the energy, and the coherences between
stationary states are zero \cite{43}.

\section{ Quantization Relationships \& Quantum Replicator Dynamics}

Let us propose the next quantization relationships%
\begin{gather}
x_{i}\rightarrow \sum_{k=1}^{n}\left\langle i\left\vert \Psi _{k}\right.
\right\rangle p_{k}\left\langle \Psi _{k}\left\vert i\right. \right\rangle
=\rho _{ii}\text{,}  \label{67} \\
(x_{i}x_{j})^{1/2}\rightarrow \sum_{k=1}^{n}\left\langle i\left\vert \Psi
_{k}\right. \right\rangle p_{k}\left\langle \Psi _{k}\left\vert j\right.
\right\rangle =\rho _{ij}\text{.}  \label{68}
\end{gather}%
It means that a population will be represented by a quantum system in which
each subpopulation playing strategy $s_{i}$ will be represented by a pure
ensemble in the state $\left\vert \Psi _{k}(t)\right\rangle $ and with
probability $p_{k}$. The probability $x_{i}$ of playing strategy $s_{i}$ or
the relative frequency of the individuals using strategy $s_{i}$ in that
population will be represented as the probability $\rho _{ii}$ of finding
each pure ensemble in the state $\left\vert i\right\rangle $. Our matrix $X$
of relative frequencies of a population corresponds to the density matrix $%
\rho $%
\begin{equation}
X\rightarrow \rho   \label{69}
\end{equation}%
and from equations (\ref{31}) and (\ref{48})%
\begin{equation}
\Lambda \rightarrow \hat{H}\text{,}  \label{70}
\end{equation}%
where $H=i\hbar \hat{H}$. Through these quantization relationships the
replicator dynamics takes the form of the equation of evolution of mixed
states%
\begin{equation}
i\hbar \frac{d\rho (t)}{dt}=\left[ H(t),\rho (t)\right]   \label{71}
\end{equation}%
and according with equation (\ref{49}) the entropy of our system would be
given by%
\begin{equation}
\sigma =-Tr(X\ln X)\text{,}  \label{72}
\end{equation}%
where $x_{ij}$ are the elements of the matrix $X.$

\bigskip

We can translate our classical definitions to the quantum world like the
payoff function%
\begin{equation}
f_{i}(\rho )=\sum_{j=1}^{n}a_{ij}\rho _{jj}  \label{73}
\end{equation}%
and the average fitness of the population%
\begin{equation}
\left\langle f(\rho )\right\rangle =\sum_{k,l=1}^{n}a_{kl}\rho _{kk}\rho
_{ll}\text{.}  \label{74}
\end{equation}%
When the non diagonal elements of $A$ and $\rho $ are zero%
\begin{equation}
Q=\frac{1}{2}A\rho   \label{75}
\end{equation}%
so that we can rewrite $\Lambda $ as%
\begin{equation}
\Lambda =\frac{1}{2}\left[ A\rho -\rho A\rho \right]   \label{76}
\end{equation}%
and due to the fact that $H=i\hbar \Lambda $%
\begin{equation}
H=\frac{i\hbar }{2}\left[ A\rho -\rho A\rho \right]   \label{77}
\end{equation}%
the average energy of our system is%
\begin{gather}
\left\langle E\right\rangle =Tr(\rho H)\text{,}  \notag \\
\left\langle E\right\rangle =\frac{i\hbar }{2}Tr\left[ \rho A\rho -\rho
A\rho \right] =0\text{.}  \label{78}
\end{gather}%
For this case it is easy to note in equation (\ref{76}) that the Hamiltonian
$H$ of our system is a diagonal matrix in where its elements $H_{ii}$
correspond to its eigenvalues and its eigenvectors are%
\begin{gather}
\left\vert \Psi _{1}\right\rangle =c_{11}\left\vert 1\right\rangle
+c_{12}\left\vert 0\right\rangle +...+c_{1i}\left\vert 0\right\rangle
+...+c_{1n}\left\vert 0\right\rangle \text{,}  \notag \\
\left\vert \Psi _{2}\right\rangle =c_{21}\left\vert 0\right\rangle
+c_{22}\left\vert 1\right\rangle +...+c_{2i}\left\vert 0\right\rangle
+...+c_{2n}\left\vert 0\right\rangle \text{,}  \notag \\
...  \label{79} \\
\left\vert \Psi n\right\rangle =c_{n1}\left\vert 0\right\rangle
+c_{n2}\left\vert 0\right\rangle +...+c_{ni}\left\vert 0\right\rangle
+...+c_{nn}\left\vert 1\right\rangle   \notag
\end{gather}%
subject to the constrain $\sum_{j=1}^{n}$ $\left\vert c_{ij}\right\vert
^{2}=1$. For the case of a two person symmetric game with $a_{ij}=0$ and $%
\rho _{ij}=0$ the eigenvectors of the game are%
\begin{gather}
\left\vert \Psi _{1}\right\rangle =\alpha \left\vert 1\right\rangle +\beta
\left\vert 0\right\rangle \text{,}  \notag \\
\left\vert \Psi _{2}\right\rangle =\gamma \left\vert 0\right\rangle +\delta
\left\vert 1\right\rangle \text{,}  \label{80}
\end{gather}%
where $\alpha ,\beta ,\gamma ,\delta $ are subject to the constrains $%
\left\vert \alpha \right\vert ^{2}+$ $\left\vert \beta \right\vert ^{2}=1$
and $\left\vert \gamma \right\vert ^{2}+$ $\left\vert \delta \right\vert
^{2}=1$. For this case also $\left[ \Lambda ,\rho \right] =\Lambda $ and%
\begin{equation}
\frac{dX}{dt}=\Lambda \text{.}  \label{81}
\end{equation}%
The payoff function takes the form $f(x)=A\rho $ and its trace is equal to%
\begin{equation}
Tr(f(x))=\left\langle A\right\rangle _{\rho }  \label{82}
\end{equation}%
and the average payoff function is given by%
\begin{equation}
\left\langle f(x)\right\rangle =Tr(\rho A\rho )\text{.}  \label{83}
\end{equation}%
For $x_{ij}=0$ our entropy takes the form%
\begin{equation}
\sigma =-\sum_{i=1}^{n}x_{ii}\ln x_{ii}=-\sum_{i=1}^{n}x_{i}\ln x_{i}\text{.}
\label{84}
\end{equation}

Finally, the equation of evolution of mixed states from quantum statistical
mechanics (\ref{48}) is the quantum analogue of \ the replicator dynamics in
matrix form (\ref{31}) and the respective matrixes of both systems have
similar properties.

\section{Quantum Strategic Equilibrium}

Quantum mechanics could be a much more general theory that we had thought.
It could encloses theories like EGT and evolutionary dynamics and we could
explain through this theory biological and economical processes. From this
point of view many of the equations, concepts and its properties defined
quantically must be more general that its classical versions which must
remain inside of the foundations of the new quantum version. So, our quantum
equilibrium concept also must be more general than the one defined
classically.

\bigskip

If in an isolated system each of its accessible states do not have the same
probability, the system is not in equilibrium. The system will vary and will
evolution in the time until it reaches the equilibrium state when the
probability of finding the system in each of the accessible states be the
same. Then the system will find its more probable configuration in which the
number of accessible states is maximum and equally probable. The whole
system will vary and rearrange its state and the states of its ensembles
with the purpose of maximize its entropy and reach its minimum energy state.
We could say that the purpose and maximum payoff of a quantum system is its
minimum energy state.

\bigskip

The system and its members will vary and rearrange themselves to reach the
best possible state for each of them which is also the best possible state
for the whole system. This can be seen like a microscopical cooperation
between quantum objects to improve its state with the purpose of reaching or
maintaining the equilibrium of the system. All the members of our quantum
system will play a game in which its maximum payoff is its minimum energy
state. Different particles will cooperate to form a system in which its new
state will have to be better than the one they had before they interact by
forming an atom. Different atoms will cooperate to form a system in which
its new state have to be better than the one they had before they formed
what is called a \textquotedblleft inorganic\textquotedblright\ molecule.

\bigskip

A system is stable only if it maximizes the welfare of the collective above
the welfare of the individual, if it is maximized the welfare of the
individual above the welfare of the collective the system gets unstable and
eventually it collapses.

\bigskip

Gafiychuk and Prykarpatsky \cite{45} applied the replicator equations
written in the form of nonlinear von Neumann equations to the study of the
general properties of the quasispecies dynamical system from the standpoint
of its evolution and stability. They developed a mathematical model of a
naturally fitted coevolving ecosystem and a theoretical study a
self-organization problem of an ensemble of interacting species. DNA is a
nonlinear dynamical system and its evolution is a sequence of chemical
reactions. An abstract DNA-type system is defined by a set of nonlinear
kinetic equations with polynomial nonlinearities that admit soliton
solutions associated with helical geometry. Aerts and Czachor \cite{46}
shown that the set of these equations allows for two different Lax
representations: They can be written as von Neumann type nonlinear systems
and they can be regarded as a compatibility condition for a
Darboux-covariant Lax pair. Organisms whose DNA evolves in a chaotic way
would be eliminated by natural selection.\ They also explained why
non-Kolmogorovian probability models occurring in soliton kinetics are
naturally associated with chemical reactions. Gogonea and Merz \cite{38}
indicated that games are being played at the quantum mechanical level in
protein folding. Turner and Chao \cite{47} studied the evolution of
competitive interactions among viruses in an RNA phage, and found that the
fitness of the phage generates a payoff matrix conforming to the two-person
prisoner's dilemma game. Bacterial infections by viruses have been presented
as classical game-like situations where nature prefers the dominant
strategies. Patel \cite{39,40} suggested quantum dynamics played a role in
the DNA replication and the optimization criteria involved in genetic
information processing. He considers the criteria involved as a task similar
to an unsorted assembly operation where the Grover's database search
algorithm fruitfully applies; given the different optimal solutions for
classical and quantum dynamics. Azhar Iqbal \cite{22} showed results in
which quantum mechanics has strong and important roles in selection of
stable solutions in a system of interacting \textquotedblleft
entities\textquotedblright . These entities can do quantum actions on
quantum states. It may simply consists of a collection of molecules and the
stability of solutions or equilibria can be affected by quantum interactions
which provides a new approach towards theories of rise of complexity in
groups of quantum interacting entities. Neuroeconomics \cite{48,49} may
provide an alternative to the classical Cartesian model of the brain and
behavior \cite{50} through a rich dialogue between theoretical neurobiology
and quantum logic \cite{51,52}.

\section{Conclusions}

It has been shown that through the quantization relationships proposed the
quantum analogue of the replicator dynamics is the equation of evolution of
mixed states of a quantum system. A population is represented by a quantum
system in which each subpopulation playing strategy $s_{i}$ is represented
by a pure ensemble in state $\left\vert \Psi _{k}\right\rangle $ and with
probability $p_{k\text{ }}$. The probability $x_{i}$ of playing strategy $%
s_{i}$ or the relative frequency of the individuals using strategy $s_{i}$
in that population will be represented as the probability $\rho _{ii}$ of
finding each pure ensemble in the state $\left\vert i\right\rangle .$
Through quantum mechanics we can describe economical and biological
processes from a different perspective and with a quantum equilibrium
concept which is more general that whichever defined classically. The
purpose and maximum payoff of a system is improve its state by maximizing
the welfare of its members through maximizing the welfare of the system. For
a quantum system the maximum payoff is its minimum energy state. The system
and all its members will cooperate and rearrange its states to improve its
present condition. A system is stable only if it maximizes the welfare of
the collective above the welfare of the individual or else the system gets
unstable and eventually it collapses.

\end{document}